\documentclass[10pt, a4paper]{article}
\usepackage{fullpage}
\usepackage{graphicx}
\usepackage[colorlinks=true, citecolor=blue]{hyperref}
\usepackage{breakurl}
\usepackage{amsmath}
\usepackage{comment}
\usepackage{amsfonts}
\usepackage{xcolor}
\usepackage{array}
\usepackage{makecell}
\usepackage{tabularx}
\usepackage{soul}
\usepackage{multirow}
\usepackage{nicematrix}
\usepackage{enumitem}
\usepackage{algorithm}
\usepackage{algpseudocode}
\usepackage[nameinlink]{cleveref}
\usepackage{float}
\usepackage[font=small,labelfont=bf]{caption}
\usepackage[
backend=biber,
style=alphabetic,
sorting=ynt
]{biblatex}
\usepackage{algorithmicx}

\title{\textbf{Atlantis Protocol}}

\author{Oleksandr Kurbatov, Kyrylo Riabov, Mykhailo Velykodnyi \\
\small v.0.1
}

\date{December 2024}

\begin{document}
\maketitle

\begin{abstract}
This document proposes a combination of several techniques to construct anonymous and untraceable payment systems.
The proposed system supports arbitrary transfer amounts and enables the simultaneous transfer of multiple assets.
\end{abstract}

\section{Introduction}
Public tokens and stablecoins are a problem for consumers. It could be quite risky if any external party can see the particular balance and trace the history of interactions and relations between users and organizations.

Several solutions solve privacy issues, but they have several technical limitations that we want to reduce. The first solution is the Mimblewimble protocol introduced in \cite{mimblewimble}. This protocol allows the creation of compact cryptographic commitments for UTXOs and efficiently aggregates them by producing only one signature value and range proofs for outputs. The drawbacks of the protocol are 1 -- the ability to track which commitments are being spent within the particular transaction and 2 -- the ability to operate only with a single currency.

The second solution we refer to is Tornado Cash \cite{web:tornado_cash_whitepaper}. This approach allows us to keep the deposit commitments within the tree and break the connections when withdrawing coins. Tornado Cash provides untraceability of payments, but at the same time, users need to operate only with the same denomination values (not allowing Sudoku analysis \cite{web:coinjoin_sudoku_advisory}) and create new contract instances to operate with different assets.

We introduce the Atlantis protocol, which allows us to inherit and combine properties from the mentioned solutions and create the payment method with the next set of features:
\begin{enumerate}
    \item \textbf{Anonymity}. No public keys or addresses represent users while transferring coins.
    \item \textbf{Untraceability}. No connection between newly created commitments and spent commitments.
    \item \textbf{Limitness}. No limitations on the amount of funds within the payment.
    \item \textbf{Heterogeneity}. It's possible to support several assets in one commitment without increasing the commitment size.
\end{enumerate}

\section{Preliminaries}

$\mathbb{G}$ a cyclic group of prime order $p$ written additively, $G \in \mathbb{G}$ is the group generator. $a \in \mathbb{F}_p$ is a scalar value and $A \in \mathbb{G}$ is a group element. $\mathsf{hash}(m) \rightarrow h\in \mathbb{F}_p$ is the cryptographic hash function that takes as an input an arbitrary message $m$ and returns the field element $h$. $\mathsf{NUMS}(i).\mathsf{toPoint} \rightarrow A \in \mathbb{G}$ is the "Nothing-up-my-sleeve" \cite{bernstein2015} operation that converts the arbitrary input to the group element $A$ the way nobody can find $a$ such as $A = aG$.

We define the relation for the proof as $\mathcal{R} = \{(w,x) \in \mathcal{W} \times \mathcal{X}: \phi_1(w,x), \phi_2(w,x) , \dots, \phi_m(w,x)\}$, where $w$ is a witness data, $x$ is a public data and $\phi_1(w,x), \phi_2(w,x) , \dots, \phi_m(w,x)$ the set of relations must be proven simultaneously.

We are referring to the Schnorr signature algorithm \cite{schnorr1989} within two functions: $\mathsf{sigGen}(sk, m) \rightarrow \mathsf{sig}$, where $\mathsf{sig} = \langle R, s\rangle$ is the signature value, $s$ is the signer's private key and $m$ is the signed message; and $\mathsf{sigVer}(\mathsf{sig}, P, m) \rightarrow \mathsf{bool}$ is a signature verification function that operates with the public key $P$ and returns the result of signature verification.

The last component is a Merkle proof $\pi_t \in \mathbb{F}_p^{(n)}$ is the list of node values that leads to the root value $t$. The Merkle proof for the element $x$ can be verified by $\mathsf{MPVer}(x, \pi_t, t) \rightarrow \mathsf{bool}$.

\section{Commitment construction}
The commitment $\mathsf{C} \in \mathbb{G}$ represents the user's balance and is constructed as follows \cite{pedersen1991}:

\[\mathsf{C} = aH + skG,\]

where:
\begin{itemize}
    \item[] $a$ -- an amount of coins in the commitment
    \item[] $sk$ -- user's secret key
    \item[] $H, G \leftarrow \mathbb{G}$ -- the set of generators
\end{itemize}

For simplification, let's say $H$ is the asset generator, and $G$ is the ownership generator. We can extend this commitment construction scheme to support several assets by:

\[\mathsf{C} = \sum_{i=0}^{n} a_iH_i + skG,\]

where $i$ is the asset identifier (it can be the contract's address of a particular asset, for example), $a_i$ is the number of coins with identifier $i$, $H_i \leftarrow \mathsf{NUMS}(i).\mathsf{toPoint}$ -- the generator that represents a particular coin type $i$.

So, $\mathsf{C}$ is a single EC point representing several user balances with appropriate coin types. Then, the main idea is that we can operate with different assets in a homomorphic way, proving the conservation rule (the number of coins in input commitments equals the number of coins in output commitments).

All commitments are stored in the Sparse Merkle Tree \cite{sparse_merkle_tree}, allowing their untraceability within transfer and withdrawal operations. At the same time, we will collect nullifiers connected to secret keys as $\mathsf{null} = \mathsf{hash}(sk)$, preventing commitment from being spent twice. We can store these nullifiers in the map on the contract or collect them in the additional tree (in these cases, users must generate exclusion proofs and add nullifiers into this tree after transfers and withdrawals).

\section{Protocol description}
We can divide the protocol into the list of operations it supports:
\begin{enumerate}
    \item \textbf{Deposit}. Allows the coins to be converted into a cryptographic commitment stored in the tree.
    \item \textbf{Transfer}. Spends the existing commitment(s) and creates a set of new unspent commitment(s).
    \item \textbf{Withdrawal}. Allows converting the commitment(s) to the public number of coins and withdrawal to the user's account.
\end{enumerate}

We can define the list of contract public parameters needed for all operations validation: $G$ is the constant ownership generator; $\langle i, H_i\rangle^n, i \in [0, n]$ -- the HashMap with contracts' identifiers and corresponding asset generators. This HashMap can be updated by adding the needed asset contract address and calculating $\langle i', H'_i \rangle \leftarrow \mathsf{NUMS}(i').\mathsf{toPoint}$ by the Atlantis contract (to be sure NUMS procedure is performed correctly).

\subsection{Deposit}
Deposit operations don't require proof; the contract can calculate the commitment based on deposit details and public values provided by the deposit initiator. At the same time, the contract must verify that the deposited asset is supported (there are no technical limitations for the list of supported contracts, but it affects the complexity of the range of proofs generated for transfer and withdrawal transactions).

\begin{algorithm}[H]
\caption{Deposit}
\label{alg:deposit}
\textbf{Inputs:}
    \begin{itemize}
        \item \textit{Private:}
        \begin{itemize}
            \item[] $sk$ -- user's secret key
        \end{itemize}
        \item \textit{Public:}
        \begin{itemize}
            \item[] $a_i, i \in [0, n]$ -- amounts of deposited coins for different asset types
            \item[] $H_i, i \in [0, n]$ -- coin types' generators
            \item[] $G$ -- ownership generator
            \item[] $P = skG$ -- deposit public key
        \end{itemize}
    \end{itemize}
\textbf{Depositing process:}
\begin{algorithmic}
    \State \qquad 1. The user makes a deposit and provides $P$ value to the contract.
    \State \qquad 2. Contract extracts $a_i$ and $H_i, \forall i \in [0,n]$ based on the deposit data.
    \State \qquad 3. Contract calculates the commitment as \begin{gather*}
            \mathsf{C} = \sum_{i=0}^{n}a_iH_i + P
        \end{gather*}
    \State \qquad 4. The contract stores the point $\mathsf{C}$ in the tree by the $\mathsf{index} = \mathsf{hash}(\mathsf{C})$
\end{algorithmic}
\end{algorithm}

While depositing funds, the user can create not one but several commitments. For that they can define $i, a_{i_{j}}$, where $j \in [0, m]$, $m$ -- is the number of commitments. Each $\mathsf{C}_j = \sum_{i=0}^{n}a_{i_{j}}H_i + P_j$ is added into the separate tree leaf.

\subsection{Transfer}
A transfer operation spends the list of existing commitments and creates new ones. However, for transfer, a sender and a recipient should cooperate in constructing the payment transaction. We can describe this process in the following steps:
\begin{enumerate}
    \item The sender calculates the list of nullifiers for commitments that need to be spent and provides it to the recipient.
    \item The recipient forms the commitment based on transfer details (amount and assets' types) and their key and then produces the signature of nullifiers. Additionally, the recipient generates a range of proof for the amount in their commitment(s).
    \item The sender produces their signatures, aggregates them with the recipient's, and generates the final proof (including commitment(s) for the change if required).
\end{enumerate}

\begin{algorithm}[H]
\caption{Transfer}
\label{alg:transfer}
\textbf{Initiation:}
\begin{algorithmic}
    \State \qquad 1. The sender has the list of unspent commitments $\mathsf{C}_s^{(n)}$ with secret keys $sk_s^{(n)}$
    \State \qquad 2. They calculate the list of nullifiers as $\forall i \in [0, n], \mathsf{null}_{s_i} \leftarrow \mathsf{hash}(sk_i)$
    \State \qquad 3. Then, the sender transfers the list $\mathsf{null}_s^{(n)}$ to the recipient
\end{algorithmic}
\vspace{0.2cm}
\textbf{Preparation:}
\begin{algorithmic}
    \State \qquad 1. The recipient generates $s_r, P_r=sk_rG$ and their output commitment as \begin{gather*}
        \mathsf{C}_r = \sum_{i=0}^{n}a_iH_i + P_r.
    \end{gather*}
    \State \qquad 2. The recipient generates the signature $\mathsf{sig}_r \leftarrow \mathsf{sigGen}(sk_r, \mathsf{null}_s^{(n)})$
    \State \qquad 3. The recipient generates the range proof $\pi_{rp_r}$ for the relation:\begin{gather*}
        \mathcal{R}_{rp} = \{a_i^{(n)}, sk_r; G, H_i^{(n)}, \mathsf{C}_r: \forall i \in [0,n], a_i \in [0, 2^{128}), \mathsf{C}_r = \sum_{i=0}^{n}a_iH_i + sk_rG\}
    \end{gather*}
    \State \qquad 4. The recipient transfers $\mathsf{sig}_r$ and $\pi_{rp_r}$ to the sender
\end{algorithmic}
\vspace{0.2cm}
\textbf{Transfer execution:}
\begin{algorithmic}
    \State \qquad 1. The sender generates the list of signatures for all input commitments \begin{gather*}
        \mathsf{sig}_{s_i} \leftarrow \mathsf{sigGen}(sk_{s_i}, \mathsf{null}_s^{(n)})
    \end{gather*}
    \State \qquad 2. The sender generates $s_c, P_c = sk_cG$ and their commitment for the change as \begin{gather*}
        \mathsf{C}_c = \sum_{i=0}^{n}a_iH_i + P_c
    \end{gather*}
    \State \qquad 3. The sender generates the signature $\mathsf{sig}_c \leftarrow \mathsf{sigGen}(sk_c, \mathsf{null}_s^{(n)})$
    \State \qquad 4. The sender generates the change's range proof $\pi_{rp_c}$ for the relation \begin{gather*}
        \mathcal{R}_{rp} = \{a_i^{(n)}, sk_c; G, H_i^{(n)}, \mathsf{C}_c: \forall i \in [0, n], a_i \in [0, 2^{128}), \mathsf{C}_c = \sum_{i=0}^{n}a_iH_i + sk_cG\}
    \end{gather*}
    \State \qquad 5. The sender calculates the aggregated signature for all input and output commitments as \begin{gather*}
        \mathsf{sig}_{agg} = \sum_{i=0}^n\mathsf{sig}_{s_i} - \mathsf{sig}_r - \mathsf{sig}_c
    \end{gather*}
    \State \qquad 6. The sender generates the final payment proof $\pi_{p}$ for the following relation \begin{gather*}
        \mathcal{R}_{p} = \{sk_s^{(n)},\mathsf{sig}_{agg},\mathsf{C}_s^{(n)}, \pi_{t}^{(n)}; \mathsf{null}_c^{(n)}, \mathsf{C}_r, \mathsf{C}_c, t: \\
        \mathsf{sigVer}(\mathsf{sig}_{agg}, \sum_{i=0}^n\mathsf{C}_{s_i} -\mathsf{C}_r-\mathsf{C}_c,\mathsf{null}_s^{(n)}) \rightarrow \mathsf{true},\\
        \forall i \in [0,n], \mathsf{null}_{s_i} \leftarrow \mathsf{hash}(sk_{s_i}),\\
        \forall i \in [0,n], \mathsf{MBVer}(\mathsf{C}_{s_i}, \pi_{t_i}, t) \rightarrow \mathsf{true}\}
    \end{gather*}
    \State \qquad 7. The sender provides $\pi_{rp_r}, \pi_{rp_c}$ and $\pi_{p}$ to the contract together with the public data. If all proofs are correct, the contract adds $\mathsf{C}_r$ and $\mathsf{C}_c$ in the tree and $\mathsf{null}_s^{(n)}$ into the list of spent commitments.
\end{algorithmic}
\end{algorithm}

In the same way, the user can send payments to several recipients within one interaction -- they need to generate the signatures and range proofs for all output commitments and provide it to the sender, which then aggregates them once.

\subsection{Withdrawal}
While withdrawing coins, the user proves there is an unspent commitment with the appropriate amount of coins with the particular types and publishes the nullifier connected to the secret key.

\begin{algorithm}[H]
\caption{Withdrawal}
\label{alg:withdraw}
\textbf{Inputs:}
    \begin{itemize}
        \item \textit{Private:}
        \begin{itemize}
            \item[] $sk$ -- user's secret key
            \item[] $\mathsf{C}$ -- user's commitment
            \item[] $\pi_t$ -- the Merkle proof for the commitment inclusion
        \end{itemize}
        \item \textit{Public:}
        \begin{itemize}
            \item[] $a_i, i \in [0, n]$ -- amounts of coins for different asset types
            \item[] $H_i, i \in [0, n]$ -- coin types' generators
            \item[] $G$ -- ownership generator
            \item[] $\mathsf{null} \leftarrow \mathsf{hash}(sk)$
        \end{itemize}
    \end{itemize}
    \textbf{Proving:}
    \begin{algorithmic}
    \State \qquad Generate proof $\pi_w$ for relation: \begin{gather*}
            \mathcal{R}_{w} = \{sk, \mathsf{C},\pi_t; a_i^{(n)}, G, H_i^{(n)}, \mathsf{null}: \mathsf{null} \leftarrow \mathsf{hash}(sk), \mathsf{MBVer}(\mathsf{C}, \pi_t, t) \rightarrow \mathsf{true}, \mathsf{C} = \sum_{i=0}^{n}a_iH_i + skG
        \end{gather*}
    \State \qquad The proof $\pi_w$ is verified by the contract. If it's correct and $\mathsf{null}$ isn't included in the list of spent nullifiers, the user withdraws coins to the public balance.
    \end{algorithmic}

\end{algorithm}

\section{Security properties and regulation aspects}
The described version of Atlantis Protocol provides the following properties:
\begin{itemize}
    \item Despite providing the user's initial public key for the deposit, it doesn't appear in transfer and withdrawal operations.
    \item When users transfer or withdraw coins, they hide the commitment in the list of all existing ones.
    \item A commitment with the same nullifier can't be spent twice (if we want to provide an ability using the same public key, we can introduce an additional "nullifier" generator and extend Pedersen's commitment with the third component).
    \item The user can operate with any amount without limitations based on particular denominations.
    \item Potentially, an infinitive number of different assets can be supported within a single commitment without increasing proof verification complexity. However, the complexity of proof generation must be taken into account.
    \item When we recommended using proofs for $[0, 2^{128})$ range, we assumed operating with 254-bit amounts (it shouldn't be possible to overflow the output number of coins with a reasonable amount of output commitments). It can be different in the final implementation.
    \item The Sudoku analysis is possible if the user makes no transfer between deposit and withdrawal operations.
\end{itemize}

The privacy-preserving properties of such a solution may raise some regulatory concerns, so we encourage using additional constructions that will help achieve \textbf{eligible privacy} and protect users from the associated risks. One is the inclusion and exclusion lists proposed in \cite{buterin2023privacy}, allowing users to select the quorum of commitments to be associated/disassociated.

If an instance of such an application supports centralized or partially decentralized governance, its administrators can forcibly add suspicious commitments to an exclusion list, making them impossible to spend.
A timelock value can be added to the deposit, up to which the commitment cannot be spent. Adding such a value slightly increases the cost of generating a proof but does not compromise the privacy of the commitment being spent (the user simply proves that the timelock in the value tree is greater than the current timestamp). If administrators don't put the commitment on the prohibited list, it can be spent after the defined locktime.

\printbibliography
\end{document}